\documentclass[12pt,twocolumn]{emulateapj}

\usepackage{epsf}
\usepackage{epsfig}

\def\1{\'\i}

\begin{document}

\title{Constraining the baryon fraction in the Warm Hot 
Intergalactic Medium at low redshifts with PLANCK data}

\author{
R. G\'enova-Santos\altaffilmark{1,2},
F. Atrio-Barandela\altaffilmark{3},
F.-S. Kitaura\altaffilmark{4}, 
J. P. M\"ucket\altaffilmark{4}}
\altaffiltext{1}{Instituto de Astrof\1sica de Canarias,
38200 La Laguna, Tenerife, Spain; rgs@iac.es}
\altaffiltext{2}{Departamento de Astrof\'{\i}sica, Universidad de La Laguna (ULL), 
38206 La Laguna, Tenerife, Spain}
\altaffiltext{3}{F{\'\i}sica Te\'orica, Universidad de Salamanca,
37008 Salamanca, Spain; atrio@usal.es}
\altaffiltext{4}{Leibniz Institut f\"ur Astrophysik,
14482 Potsdam, Germany; kitaura,jpmuecket@aip.de}

\begin{abstract}
We cross-correlate foreground cleaned Planck Nominal Cosmic Microwave
Background (CMB) maps with two templates constructed from the Two-Micron All-Sky Redshift Survey of
galaxies. The first template traces the large-scale filamentary distribution characteristic
of the Warm-Hot Intergalactic Medium (WHIM) out to $\sim 90h^{-1}$Mpc. The second 
traces preferentially the virialized gas in unresolved halos around galaxies. 
We find a marginal signal from the
correlation of Planck data and the WHIM template with a signal-to-noise
from $0.84$ to $1.39$ at the different Planck frequencies, and with
a frequency dependence compatible with the thermal Sunyaev-Zel'dovich (tSZ)
effect. When we restrict our analysis to the 60\% of the sky outside
the plane of the Galaxy and known point sources and galaxy clusters, the
cross-correlation at zero lag is $0.064\pm 0.051\mu$K. 
The correlation extends out to $\approx 6^\circ$, which at the median depth
of our template corresponds to a physical length of $\sim 6-8~h^{-1}$Mpc.
On the same fraction of the sky, the cross-correlation of the CMB data
with the second template is $<0.17~\mu$K (95\% C.L.), 
providing no statistically significant evidence of 
a contribution from bound gas to the previous result. 
This limit translates into a physical constraint on the properties of the
shock-heated WHIM of a log-normal model describing the weakly nonlinear
density field. We find that our upper limit is compatible with
a fraction of 45\% of all baryons residing in filaments at overdensities
$\sim 1-100$ and with temperatures in the range $10^{4.5}-10^{7.5}$K,
in agreement with the detection at redshift $z\sim 0.5$ of \citet{van_waerbeke14}.
\end{abstract}

\keywords{Cosmic Background Radiation. Cosmology: theory. Cosmology: observations}

\section{Introduction}

The cosmological evolution of baryons within dark matter halos is 
still an open problem.  While at redshifts $z\sim 2$ the baryon fraction 
residing in collapsed structures is close to the cosmic mean
\citep{rauch97,weinberg97}, at redshifts $z\sim 0$ is $\sim 40-50$\% 
smaller than the measured value \citep{fukugita04, shull12}. Early hydrodynamical 
simulations suggested that the unidentified baryons could reside in 
mildly-nonlinear filamentary structures with temperatures $0.01-1$~keV 
and overdensities $\delta_{\rm B}<10^3$ the mean baryonic density, 
known as Warm-Hot Intergalactic Medium (WHIM) \citep{cen99,dave01}.
Understanding the distribution of baryons and dark matter from galaxies 
to clusters is essential to explain the formation and evolution of structure 
in the Universe. Consequently, a great observational effort is being devoted 
to detect the WHIM signature, mostly by searching for absorption lines due 
to highly ionized heavy elements. Some detections have been reported in the 
far-ultraviolet \citep {tripp08} and in the soft X-ray \citep{nicastro05,buote09} 
bands but these detections are ambiguous or controversial. In some 
cases they could not be confirmed by later studies \citep{yao12}. In others
it has not yet been settled whether the detected lines arise 
from the WHIM or from individual galaxies or groups within the filaments 
\citep{williams13}. Attempts have also been made to detect signatures 
of the WHIM diffuse X-ray emission; \citet{werner08} reported a detection from 
a filament connecting two clusters, with properties compatible with the WHIM.

The baryons in this WHIM plasma do not only contribute to the diffuse 
X-ray background but they also produce CMB distortions by means of the 
thermal Sunyaev-Zeldvich (tSZ) effect \citep{sunyaev70}. This effect could 
probe the physical state of baryons in the WHIM phase better than X-rays 
thanks to its milder dependence on the gas density. An alternative to 
finding a direct evidence towards superclusters of galaxies \citep{molnar98,genova05} 
is to correlate CMB maps with galaxy templates tracing large scale structures. 
This approach was first used by \citet{hernandez04}, whose measured signal 
was dominated by clusters of galaxies. A similar analysis was carried out by
\citet{suarez13a} and, again, the correlation was most-likely 
due to foreground residuals and to random alignments between 
structure in the data and the matter template, and not to WHIM. 
In \citet{atrio06} and \citet{atrio08} we computed the power spectra 
of the thermal and kinetic SZ anisotropies due to the WHIM, calculations 
that were later refined by \citet{suarez13b}. We used these models in 
\citet{genova09,genova13} to fit these SZ spectra to CMB data 
through a Monte-Carlo Markov Chain analysis. Although adding the SZ from 
the WHIM to the $\Lambda$CDM CMB power spectrum improved the quality 
of the fit, this improvement was not statistically significant and it did not
favor the inclusion of a new component in the model. 

While these earlier attempts based on WMAP data only provided upper 
limits to the WHIM contribution, the tSZ effect has been shown to be effective 
at detecting baryonic filaments in the denser and hotter medium of interacting 
cluster pairs \citep{pip8}. This study was greatly benefited from the finer 
angular resolution and wider frequency coverage of {\it Planck} data, 
which helped to isolate better the tSZ signal from other components. Recently, 
\citet{van_waerbeke14} reported a positive correlation
between mass maps reconstructed from the Canada France Hawaii 
Telescope Lensing Survey and a tSZ map constructed from Planck data 
with an overall significance of $6\sigma$ and extending out to 
linear scales of $\sim 10$~Mpc. They interpreted this correlation as arising 
from WHIM at $z\sim 0.4$ and with temperatures in the range $10^5-10^6$~K,
depending on the gas bias and the electron density on the filaments. However, 
the measured signal could also come from gas within virialized halos,
especially after the \cite{pip11} showed that the tSZ anisotropy exists 
down to masses as low as $\sim 2\times 10^{11}~M_\odot$.
\citet{ursino14} suggested to cross-correlate X-ray and SZ maps to 
trace the WHIM distribution and found that the small-scale
correlation would be dominated by the high-redshift WHIM, that
again would probe the WHIM at early times. 

The \citet{van_waerbeke14} detection could indicate that a large 
fraction of the ``missing'' baryons reside in a low-density warm 
plasma that traces the dark matter distribution. Equally important 
to this high redshift observation is to probe the state of the WHIM gas at 
lower redshifts. If filaments reside in the large-scale dark matter potential 
wells, then tracers of dark matter distribution at low redshifts will correlate 
with the WHIM signal both in X-ray and radio, irrespectively of the redshift.
If the gas is confined to low mass halos then its tSZ
will correlate with the distribution of galaxies. In this article we compute 
the cross-correlation of the temperature fluctuations of the CMB 
with two templates: (A) the projected matter 
density reconstructed from the distribution of galaxies within 
$\le 150h^{-1}$Mpc from the Local Group  \citep{kitaura12} to investigate 
the properties of the WHIM at redshifts $z<0.05$ and (B) the density of 
the same galaxies projected along the line of sight to study the contribution
of virialized gas in halos below the mass threshold of the matter template. 
The outline of the paper is as follows: 
In Section 2 we describe our data sets; we describe 
our procedure for reducing the foreground contribution of Planck 
Nominal maps corresponding to the March 2013 release \citep{cpp1} and 
the construction of the galaxy and the dark matter templates.
In Section 3 we explain the filtering and the masking 
that we apply to our data. In Section 4 we present the results 
of the correlation between the templates and the CMB data; in Section 5 we
discuss the implications on the WHIM physical models and, finally,
in Section 6 we summarize our main conclusions.

\section{The Data}

\subsection{Planck foreground cleaned maps}\label{sec:clean_maps}

As a tracer of the tSZ signal we use Planck DR1 maps \citep{cpp1}, 
which cover a wide frequency range delimited by nine independent channels, 
provided by the LFI (30, 44 and 70~GHz) and the HFI (100, 143, 
217, 353, 545 and 857~GHz) instruments.
Apart from the cosmological signal, and the potential tSZ signature, 
these maps contain significant foreground contamination, mainly synchrotron and 
free-free emissions at low frequency and thermal dust at high frequency, as well 
as CO emission and zodiacal light. We carefully clean all these contributions 
to obtain foreground cleaned maps at each frequency that will be a combination 
of three components: CMB, tSZ and instrumental noise. At the HFI frequencies, where 
the zodiacal light is more important, we use publicly-available maps where 
this contaminant has been removed, as described in \citet{cpp6}. All these maps are 
produced using HealPix\footnote{\tt http://healpix.jpl.nasa.gov/} 
\citep{healpix} pixelization, with resolution of $N_{\rm side}=1024$ 
and 2048 for the LFI and HFI instruments, respectively. All 
Planck-related products used in this paper have been downloaded 
from the Planck Legacy Archive\footnote{\tt 
http://www.sciops.esa.int/index.php?project=planck\&page\\
=Planck\_Legacy\_Archive}.

To perform the correction of the thermal dust emission we use the maps 
from the Planck dust model that, in each sky pixel with HealPix 
resolution $N_{\rm side}=2048$, contain the three parameters 
that define the modified black-body (MBB) emission law (dust-grains temperature, 
emissivity index and optical depth) at the reference frequency of 353~GHz. 
These maps have been obtained on a pixel-by-pixel basis by a 
$\chi^2$-minimization fit of the model to the HFI and IRAS data, 
as explained in \citet{cpp11}. In order to accurately estimate the 
contribution of this emission at each Planck frequency we evaluate the 
spectral model in each pixel and convolve it with the passband of 
each detector. We apply this color correction using the publicly-available 
routine {\tt hfi\_colour\_correction}. Similarly, to clean the synchrotron 
and free-free emissions we use maps, delivered by the Planck Collaboration,
that at each pixel contain the amplitude of the signal at the reference 
frequency of 30~GHz and a spectral index obtained by fitting 
a power-law model to the Planck lowest frequencies \citep{pes}. 
In this case a color correction using the same routine mentioned above, 
is also applied to estimate the flux weighted in each band. The maps of the 
low-frequency and high-frequency foregrounds are subtracted from 
each frequency map. 

The last step of the process is to clean the CO emission, which is 
especially important in the 100, 217 and 353~GHz channels as a result of 
the (1-0), (2-1) and (3-2) rotational transition lines, respectively. As 
explained in the \citet{pes} and in \citet{cpp13}, the Planck Consortium 
has delivered three different types of CO correction maps. The type 2 maps 
have the highest signal to noise, at the expense of having possible 
residuals from other foregrounds, but they are available
only for 100 and 217~GHz. Type 1 maps are available for the three 
frequencies, but have a much lower signal-to-noise and we found them
not to be useful for cleaning the data. Then, we do not apply any 
correction to 353~GHz and  we clean the 100 and 217 GHz maps using the 
type 2 maps, a procedure that noticeably reduces the CO contamination 
at these frequencies.

\begin{figure}[t]
\centering
\plotone{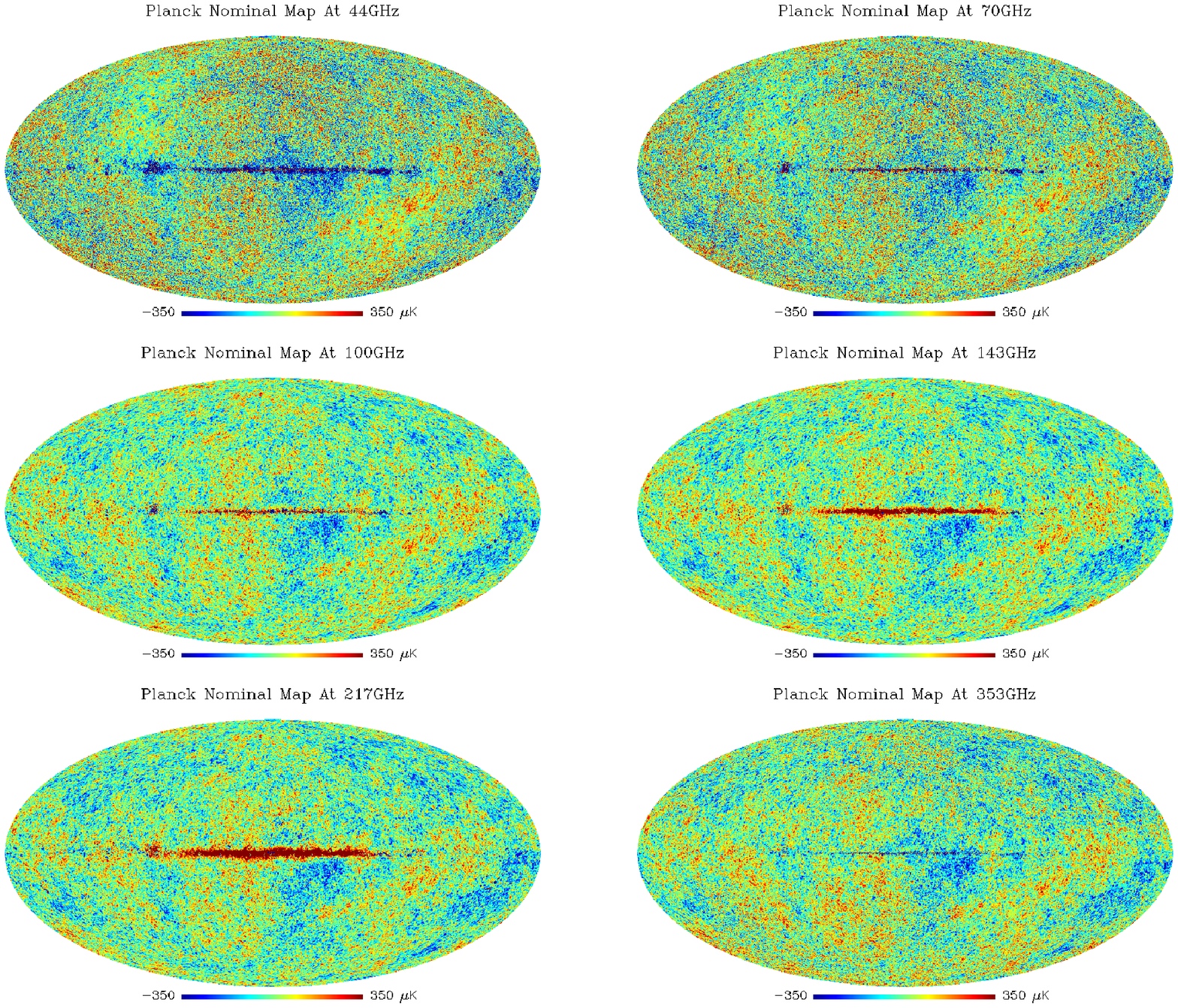}
\caption{\small 
Foreground cleaned Planck Nominal maps at different frequencies. 
From top to bottom and left to right, frequencies are 44, 70, 100, 143, 217 
and 353~GHz. A detailed explanation of the cleaning methodology is given in 
section~\ref{sec:clean_maps}. Apart from some small foreground residuals along the 
Galactic plane (which are positive in the case of the thermal dust emission and 
negative for the synchrotron emission), these maps are dominated 
by primordial CMB.
}
\label{fig1}
\end{figure}

Our final foreground cleaned maps are shown in Figure~\ref{fig1}, for frequencies 
44 to 353~GHz. The foreground contamination in these maps has been clearly 
reduced. Although there are some residuals of Galactic emission along the 
plane, the signal in the rest of the sky is dominated by the CMB fluctuations. 
The power spectra of these maps nicely match the theoretical spectrum of
the concordance model with the cosmological parameters measured by Planck.
The foreground cleaning procedure has operated best at 353 GHz since is 
at this frequency where the Planck dust model was evaluated. The cleaning at
other frequencies is affected by possible errors in the MBB 
model or in the determination of the emissivity index. The residual dust 
contamination in the Galactic plane diminishes at lower frequencies. At 
44 and 70 GHz some residual from synchrotron and free-free emission 
are visible along the plane. Although not shown here, the 30~GHz map was also used 
in our analysis. On the contrary, we ignored the 545 and the 857~GHz due to 
having stronger foreground residuals.

In order to avoid sky regions affected by emission of diffuse foregrounds, 
point sources or galaxy clusters, we apply different masks. 
We first apply the Planck mask derived from the Commander-Ruler component separation 
method\footnote{Mask named {\tt COM\_CompMap\_Mask-rulerminimal\_2048\_R1.00.fits} 
in the Planck Legacy Archive.} \citep{cpp12}, 
which retains $82\%$ of the sky. Most of the masked pixels are associated 
with point sources; a stripe along the Galactic plane of $\sim 10^\circ$
width in Galactic latitude is also removed, accounting for about $8\%$ 
of the sky. This Galactic mask is sufficient to remove the residual Galactic 
emission that appear in the maps of Figure~\ref{fig1}. To avoid regions affected 
by potentially fainter residuals we apply more conservative masks. 
We perform the analysis using three different Planck Galactic masks\footnote{To 
be found in the {\tt HFI\_Mask\_GalPlane\_2048\_R1.10.fits} file.}.
The total sky fractions retained in the three resulting masks are
respectively $73.3$, $55.8$ and $37.1\%$, but we follow the practice to
refer to these masks as the $70\%$, $60\%$ and $40\%$ masks.
Finally, we need to eliminate possible tSZ signals from Galaxy clusters. 
Although the Commander-Ruler mask already removes some clusters, to be 
conservative we also mask all pixels within a radius of $0.5^\circ$ around 
all known clusters with redshifts $z\le 0.04$ and with X-ray luminosities 
$L_{\rm X}\ge 10^{43}$erg/s  in the ROSAT [0.1-2.4]KeV band. 
The cross-correlation functions are computed
on maps with a resolution $N_{\rm side}=128$. All masks are downgraded to
this pixel resolution and, conservatively, we only retain pixels with 
values greater than $0.75$.

\subsection{Density field templates}\label{sec:templates}

We will analyze two possible distributions of the missing baryon
fraction: (A) unbound gas following the filamentary structure of the
large scale matter distribution, and (B) virialized gas in unresolved
halos traced by the spatial distribution of galaxies. 
Hereafter, we will refer to the two templates tracing these
distributions as the matter template $M$ and the galaxy template 
$M_{\rm g}$, respectively. The reconstruction 
technique of the matter density field is based on a Bayesian 
Networks Machine Learning algorithm (the Kigen-code) which self-consistently 
samples the initial density fluctuations compatible with the observed galaxy 
distribution and a structure formation model given by second order 
Lagrangian perturbation theory \citep{kitaura12,kitaura12b}. 
As a tracer of the  matter density field we used the Two-Micron All-Sky 
Redshift Survey (2MRS, Huchra et al. 2012). The method recovers 
the non-linear structures like knots, sheets, filaments and voids 
in the cosmic web to great accuracy. 
The reconstruction is performed on a cubic box 
since it requires to compute FFTs. In previous works we reconstructed 
the density field using boxes of side $160h^{-1}$Mpc \citep{kitaura12b} and 
$180h^{-1}$Mpc \citep{hess13} and found compatible results, 
proving that the boundary effects are unimportant. In this work, the
reconstructed density field on the sphere includes modes
with wavelengths up to $110-130h^{-1}$Mpc. For our theoretical estimates
we take the average depth of the reconstructed density field to be
$90h^{-1}$Mpc. The template built through this reconstruction technique is 
produced using a HealPix resolution $N_{\rm side}=128$.
The matter distribution shows well defined filaments, characteristic of
the mildly non-linear regime that we assume to be a good tracer of the
WHIM. As the reconstruction also traces the galaxy distribution, 
it is sensitive to the hot halos surrounding galaxies that have been 
recently detected \citep{anderson11,dai12}.

Low-mass galaxy clusters or galaxy groups below the threshold mass
of our previous template could also produce a detectable 
tSZ anisotropy \citep{pip11}. The pixels at $N_{\rm side}=128$ have an angular size of 
$\sim 0.5^\circ\times 0.5^\circ$ and if the virialized and unbound gas were
to follow the same spatial distribution we would not have 
enough angular resolution to disentangle a potential signal associated 
to collapsed low-mass systems from the signal of the unbound WHIM gas. 
Most likely the virialized gas detected by {\it Planck}
is better traced by the galaxies themselves. Under this assumption, we
can check the amplitude of this contribution using a template of the projected 
density of galaxies. To make the comparison more meaningful, we project 
along the line of sight the same sample of 2MRS galaxies that went into the 
reconstruction of the matter density field described above.
By construction, this galaxy template traces the hot circumgalactic
halo. At 10$h^{-1}$Mpc a pixel of half degree side 
corresponds to $80h^{-1}$Kpc, slightly larger than the halo 
but at $60h^{-1}$Mpc, the average depth of our template, it corresponds
to $\sim 500h^{-1}$Kpc, much larger than the virial radius of the detected 
halos \citep{anderson11}. Therefore, this template 
could potentially also trace the gas outside the circumgalactic halo.

\section{Data Processing}\label{sec:data_processing}

The tSZ signal due to WHIM has an amplitude of about few $\mu$K
but the expected cross-correlation with a matter template 
out to $\sim 90-120h^{-1}$Mpc is expected to be 
$\sim 0.03-0.3~\mu$K \citep{suarez13a}. At any frequency $\nu$, the 
error bar on the cross-correlation of the CMB data $T(\nu)$ with a matter 
template $M$, $\langle T(\nu) M\rangle(\theta)$, is dominated by the 
intrinsic CMB anisotropy. In order to obtain a statistically significant
detection, the data needs to be processed further. There are two possibilities:
(1) use the 217~GHz channel, the frequency of the tSZ null, 
to subtract off the intrinsic CMB signal while preserving the tSZ anisotropy
and (2) construct a Compton parameter $y$-map combining all frequencies
to remove all components from the data except the tSZ, as it was done in \citet{cpp21}.
Subtracting the 217~GHz channel from the other maps will increase
the noise and add the 217~GHz foreground residuals to those remaining at
other frequencies but, when constructing a $y$-map the frequency information
of the tSZ effect will be lost. This information could be crucial to distinguish
correlations due to random alignments or stripes present in the data
from the true effect due to the WHIM. Therefore, in this article we
will use method (1), i.e. we will subtract the correlation function at 217~GHz 
from those at other frequencies. 

\begin{figure}[t]
\centering
\includegraphics[width=12cm]{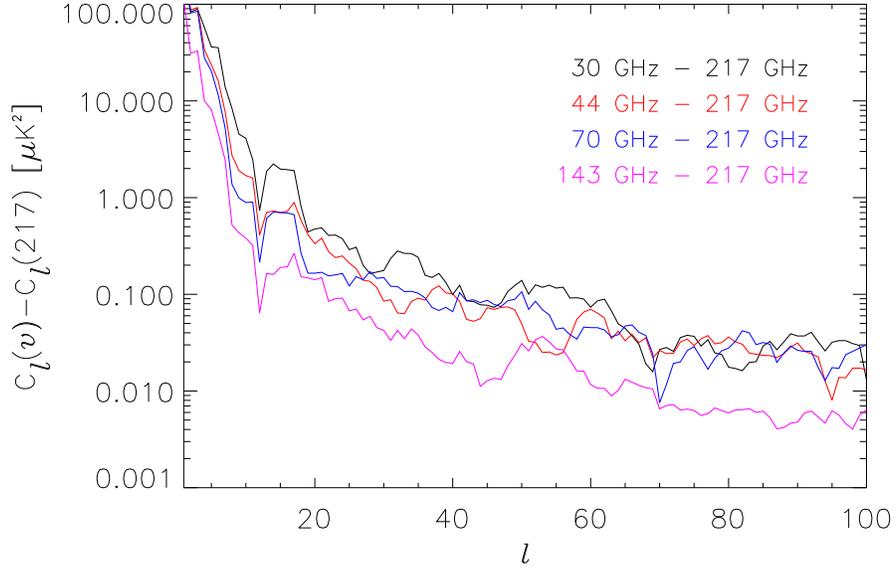}
\caption{\small 
Power spectra of the foreground cleaned Planck Nominal maps, after 
dividing by the window function of each frequency and subtracting the 
217~GHz channel. The increase in power at low multipoles could be 
produced by a combination of detector $1/f$ correlated noise and 
foreground residuals. To remove this effect we filter the cleaned maps 
by removing all multipoles with $\ell < 20$, as explained in 
section~\ref{sec:data_processing}.
}
\label{fig2}
\end{figure}

In Figure~\ref{fig2} we plot the power spectra of the subtracted maps. 
We only analyzed maps with HealPix resolution $N_{\rm side}=128$.
Effectively, all the maps have been downgraded to the same resolution by 
the pixel window, which dominates over that of the beam. When subtracting the 
217~GHz from the maps at other frequencies, we are effectively removing 
the intrinsic CMB signal. At multipoles $\ell\le 20$, the 
power spectra scale as $C_\ell\sim \ell^{-2}$, compatible with being 
dominated by marginal CMB residuals, instrumental 
$(1/f)$ noise and foreground residuals, while at $\ell\ge 60$ the spectra are 
compatible with white noise. Maps could be correlated even if there is 
not any WHIM signal present. To reduce these spurious correlations, 
maps can be filtered by removing the lowest multipoles up to any given 
$\ell_{\rm cut}$. Since at $N_{\rm side}=128$ the highest multipole is
$\ell_{\rm max}\sim 256$, removing modes up to $\ell_{\rm cut}=60$ 
could erase a significant part of the WHIM signal. As a compromise, 
we set $\ell_{\rm cut}=20$. Finally, to reduce the transfer of power from 
galactic residuals to multipoles at all scales, before computing the multipole 
expansion of the original map, 20\% of the sky around the galactic plane 
is masked. 

\begin{figure}[t]
\centering
\plotone{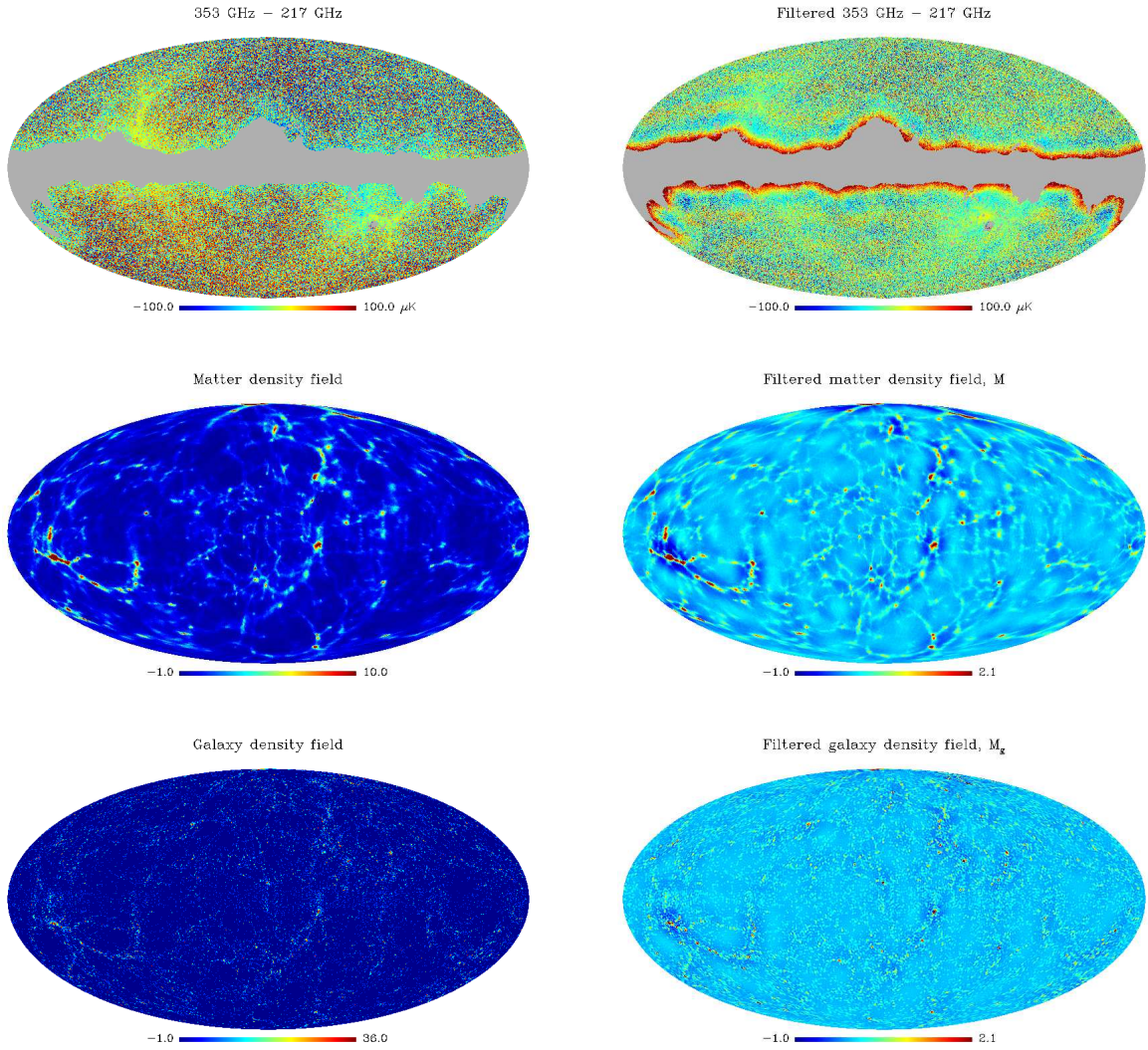}
\caption{\small 
Comparison of the original CMB data and templates with their
filtered counterparts obtained by removing all multipoles with $\ell<20$.
Top row: original Planck Nominal map at 353~GHz minus 217~GHz (left) and
filtered map (right). The mask applied to the data removes $20\%$ of 
the sky. Middle row: original (left) and filtered 
(right) matter template. Bottom row: original template
of projected density of galaxies (left) and filtered (right) template.
}
\label{fig3}
\end{figure}

In Figure~\ref{fig3} we show the 353-217~GHz difference CMB map (top row),
the matter template (middle) and the galaxy template (bottom) before and 
after the lowest multipoles have been removed. If initially the
CMB data is dominated by noise and foreground residuals, cutting the
lowest multipole reduces both contributions. Stripes
due to {\it Planck} scanning strategy are still present and, with the 
smaller noise levels, they are more clearly seen in the filtered map. 
Filtering introduces aliasing between different
multipoles, giving rise to significant temperature
anisotropies next to the galactic plane; in addition, fringing is also
present. Nevertheless, these artifacts do not pose a serious problem. In fact,
we can compute the CMB-template cross-correlation with different masks, 
removing from 20\% to 60\% of the sky, to test the effect of these systematics.

For consistency we also remove the lowest multipoles of the templates. The 
matter template is shown in the middle row of Figure~\ref{fig3} before (left
panel) and after (right panel) filtering. Since the template has little
power at large scales (see Figure~2 of Suarez-Vel\'asquez 2013a),
the filtered template has very similar structure to the original
template and the filaments show a very similar distribution.
The template of the projected density of galaxies
before (left) and after (right) filtering is shown in the bottom row 
of Figure~\ref{fig3}. This template has a similar large structure 
as the matter distribution but is less smooth at small scales as 
one could expect of the spatial distribution of clustered halos. 
Again, notice that filtering the lowest multipoles does not
affect the overall distribution of galaxies.

A visual comparison of the filtered templates of
matter (right middle panel of Figure~\ref{fig3}) and
projected density of galaxies (right bottom panel) shows
that they have a very similar large scale structure, as one could
expect since the galaxy template contains all galaxies that were
used to reconstruct the matter distribution. Their main
difference is at small scales where the matter template is a
smoother function. Then, one can assume that the matter template 
would trace better the unbound WHIM gas residing in the large scale
dark matter potential wells while the galaxy template would trace
the circumgalactic gas in virialized halos.  However, as discussed
in Sec.~2.2, both templates could trace both gas components,
probably with different relative weights, since the matter template includes 
the contribution from the hot halos around galaxies and for most
galaxies the angular scale subtended by the virial radius of those
circumgalactic halos is smaller than the pixel size at 
Healpix $N_{\rm side}=128$ resolution.

To summarize, our pipeline performs the following steps on the foreground
cleaned Planck Nominal maps: first, the galactic plane, point sources and 
clusters are masked out. Second, CMB data and templates are transformed to 
spherical harmonic space and all multipoles with $\ell<20$ are removed. 
Next, all maps are downgraded to a HealPix resolution of $N_{\rm side}=128$, 
the resolution of the matter template. Templates are normalized to zero 
mean and unit variance so the cross-correlation is given in temperature 
units. Finally, the cross-correlations $C_\nu=\langle T_\nu*M \rangle$ 
are calculated at each frequency $\nu$ using three different masks that 
preserve 70, 60 and 40 per cent of the sky. 

\section{Results}

\begin{figure}[t]
\centering
\plotone{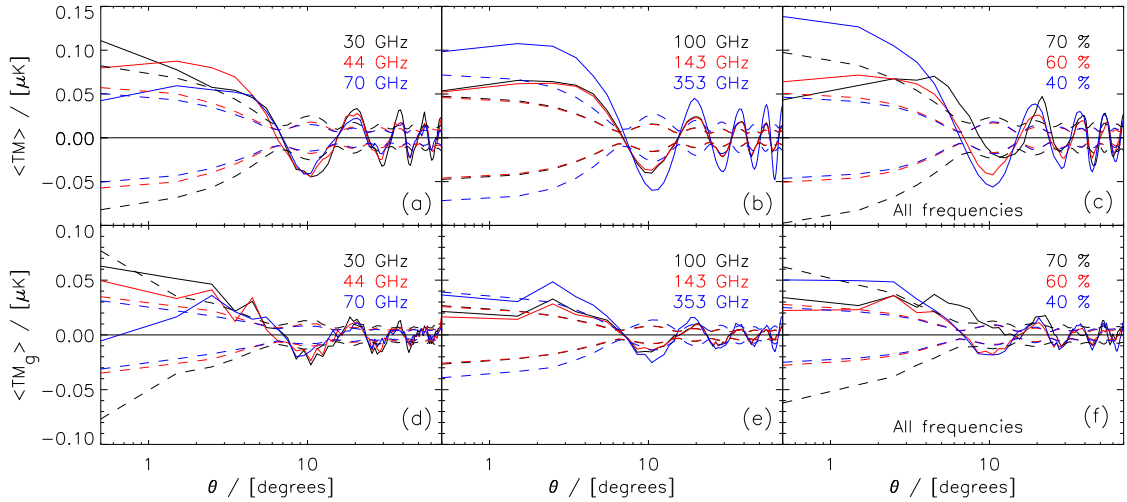}
\caption{\small 
Cross-correlation of filtered matter (top row) and galaxy density (bottom row)
templates with the foreground cleaned Planck Nominal maps divided by the 
frequency dependence of the tSZ effect; at the different frequencies
the cosmological signal was removed by subtracting the 217~GHz channel. 
Solid lines represent the data and dashed lines the error bars.  
In (a,b) and (d,e) we plot the correlations with the LFI and the HFI 
channels, respectively, calculated using the mask that retains $60\%$ of 
the sky. In (c,f) we plot the average correlation of all frequencies
obtained after masking different fractions of the sky.
The percentage indicates the fraction of the sky used.
}
\label{fig4}
\end{figure}

In order to remove the contamination from the primordial CMB, we subtract the 
217~GHz correlation function from those calculated at all other frequencies. 
If the cross-correlation $(C_\nu-C_{\rm 217})$ is dominated
by the tSZ effect due to the WHIM traced by the template, then
it must scale as $(C_\nu-C_{217})\sim G(\nu)$, reflecting
the frequency dependence of the tSZ effect $G(\nu)$. In this case,
the reduced cross-correlation $\langle TM\rangle =(C_\nu-C_{217})/G(\nu)$
between a template and the CMB data must 
be independent of frequency and proportional to the cross-correlation 
of the template with the Comptonization parameter $Y_c$,
i.e., $\langle TM\rangle\propto \langle Y_cM\rangle$. This reduced correlation
function is given in  Figure~\ref{fig4}; the top row corresponds
to the correlation with the matter template (Figure~\ref{fig3} middle right)
and the projected galaxy density template (Figure~\ref{fig3} bottom right).
Panels (a), (d) correspond to the correlation with LFI frequency maps and (b), (e) to
the correlation with HFI frequency maps computed on 60\% of the sky; (c) and (f) is 
the combination of all frequencies for different fractions of the sky.
The tSZ is negative in the Raleigh-Jeans part of the 
spectrum and is positive in the Wien region. 
The frequency dependence of the tSZ effect varies
from $G$(30~GHz)$\simeq -2$ to G(143~GHz)$\simeq -1$ and 
G(353~GHz)$\simeq 2.2$. Then, if the cross-correlation
of the templates and Planck foreground cleaned data corresponds to a
tSZ contribution, it must be negative for the frequencies below 217~GHz 
and positive above it. This is what it actually happens. Panels (a), (b), 
(d) and (e) of Figure~\ref{fig4} show that all correlations have the same sign 
once divided by $G(\nu)$, indicating that the they are consistent with 
being due to tSZ contribution since their amplitude is similar in all channels and 
have the correct sign and, although none of our measurements 
is statistically significant, the correlations with the matter template in the top 
row are a factor three larger than those with the projected galaxy density in the 
bottom row. At the origin, the signal-to-noise ratio (SNR) is of order unity, 
being somewhat larger at angular scales $\theta\sim 2-5^\circ$. 

In Figure~\ref{fig4} error bars were computed from the rms 
dispersion of the correlation of 100 random rotated templates with 
the CMB maps.  To explore the sky homogeneously, we select a random set 
of $N_{\rm sim}$ pixels drawn from a random distribution of all the pixels in the
map and compute their angular coordinates $(l,b)$. Then, we compute the Euler 
angles that align the north galactic pole with the direction $(l,b)$ 
of the randomly chosen $N_{\rm sim}$ pixels and rotate the template.
Since our template reconstructs the full sky, the area 
on which the correlation is computed is always defined by the 
mask, and therefore is the same in all the rotations. By using the same 
data to compute the correlation and its error bars we include all contributions 
due to foreground residuals and $1/f$ instrumental noise. 

In panels (c) and (f) of Figure~\ref{fig4}, we plot the correlation functions 
after combining the six LFI+HFI channels. For comparison, we plot the results 
for the three masks used in our study.  To combine different frequencies, for 
each angle we make a weighted average of the correlation functions for 
different frequencies:
\begin{equation}
\langle TM\rangle_{\rm av}(\theta) = \frac{\sum_i \langle TM\rangle (\theta,\nu_i) 
w(\theta,\nu_i)}{\sum_i w(\theta,\nu_i)}~~,
\end{equation}
where the sum runs over all the frequencies considered, and 
$w(\theta,\nu_i)=1/\sigma(\theta,\nu_i)^2$, with $\sigma(\theta,\nu_i)$ 
the rms dispersion calculated at position $\theta$ in all the rotations 
of map $\nu_i$. We calculate the error bar associated to this 
estimator by accounting for possible correlations between frequencies:
\begin{equation}
\sigma (\langle TM\rangle_{\rm av}(\theta)) = 
\frac{1}{\sqrt{\sum_{i}\sum_{j}w(\theta,\nu_i) \langle TM\rangle_{ij}
(\theta)w(\theta,\nu_j)}}~~,
\end{equation}
where 
$\langle TM\rangle_{ij}(\theta)=\langle \langle TM\rangle(\theta,\nu_i)
\langle TM\rangle(\theta,\nu_j) \rangle$ represents the covariance between 
frequencies $\nu_i$ and $\nu_j$ at position $\theta$.
We found that the results for difference frequencies are strongly correlated and
for this reason the statistical significance does not improve much by averaging 
different frequencies. In particular, at zero lag and for the $60\%$ mask, where
the covariance is typically between 0.75 and 0.85, we obtain a SNR$=1.26$, 
while the signals for individual frequencies have SNR between $0.84$ and $1.39$.
Our overall statistical significance is never larger than SNR$\simeq 2$.
After averaging over all frequencies,
the correlation with the matter template, measured on 60\% of
the sky, was $\langle TM\rangle(0)=0.064\pm 0.051~\mu$K.
Taking this result as an upper limit we obtain $\langle TM\rangle(0)< 0.11~\mu$K 
at $68\%$ C.L. and $<0.17~\mu$K at $95\%$ C.L.  Even if the result is 
not statistically significant, we find that the cross-correlation is zero at 
$\theta\sim 6-8$ degrees. The average depth of our template is $\sim 60h^{-1}$Mpc.
At this depth, the zero crossing of the correlation function corresponds to
a linear scale of $\sim 6-8~h^{-1}$Mpc. This would be the averaged projected
linear size of the WHIM filaments traced by our matter template.
This size is reassuringly consistent with the scale found by 
\citet{van_waerbeke14} and could indicate that the effect is real but
our template is not deep enough to produce a significant detection.

Comparison of the cross-correlation using different galactic masks allow
us to test the effect of fringing and power leakage of the galactic
foregrounds into the galactic poles due to our filtering scheme.
The oscillation pattern at large distances of the correlation functions shown 
in Figure~\ref{fig4} is partly produced by this residual fringing.  In addition,
the number of pairs varies with a scale of $\sim 6^\circ-10^\circ$,
the mean separation between filaments and also the typical angular
size of filaments, increasing and decreasing the sample 
variance component of the error bar accordingly. 
Notice that when we restrict the analysis to the cleanest regions 
of the sky, located around the Galactic Poles,
the statistical significance of our results increases because
even if the error bars are larger due to a bigger
sampling variance the signal increases even more. 

If the galaxy template traces preferentially the gas stored in halos,
it can be used to constrain the contribution due to the halos
below the mass threshold of the galaxy clusters excised from our analysis.
The cross-correlation between the galaxy density template and 
the foreground cleaned maps at zero lag in 60\% of the sky is also compatible with 
zero: $\langle TM_g\rangle(0)=0.022\pm 0.028~\mu$K. We obtain
an upper limit of $\langle TM_g\rangle(0)<0.078~\mu$K at the 95\% 
confidence level. As we have argued, the matter and the galaxy template
trace both the bound and unbound gas components, probably weighing differently
each contribution. Comparing with the previous result,
the galaxy template provides a more restrictive upper limit that could
reflect that the template does not trace the WHIM fully.

\section{Constraints on WHIM model parameters}

We can use the 95\% upper limits given in the previous section
to constrain the physical state of the gas in filaments. In \cite{suarez13b} we 
described the distribution of baryons in the WHIM as a network of filaments
following a log-normal distribution function.  In our model baryons
follow the dark matter distribution except at small scales, where
baryon perturbations are damped by shock heating \citep{klar10}.
At any given redshift, the cut-off scale $L_{\rm cut}$ is 
determined by the condition that the linear velocity perturbation 
${\bf v}(\bf x,z)$ is equal to or larger than the sound speed 
$c_{\rm s}=(k_{\rm B}T_{\rm IGM}(z)/m_{\rm p})^{1/2}$, with $m_{\rm p}$ 
the proton mass and $k_{\rm B}$ the Boltzmann constant. Then, the cut-off
scale $L_{\rm cut}$ can be parametrized in terms of the mean
Intergalactic Medium temperature, $T_{\rm IGM}$. While the latter 
changes little with redshift \citep{tittley07}, $L_{\rm cut}\approx 
L_0(1+z)^{1/2}$. At $T_{\rm IGM}=10^4$~K, the cut-off scale today 
would be $L_0=1.7h^{-1}$~Mpc (for details see Suarez-Vel\'asquez, 
M\"ucket \& Atrio-Barandela, 2013b).

\begin{figure}[t]
\centering
\plotone{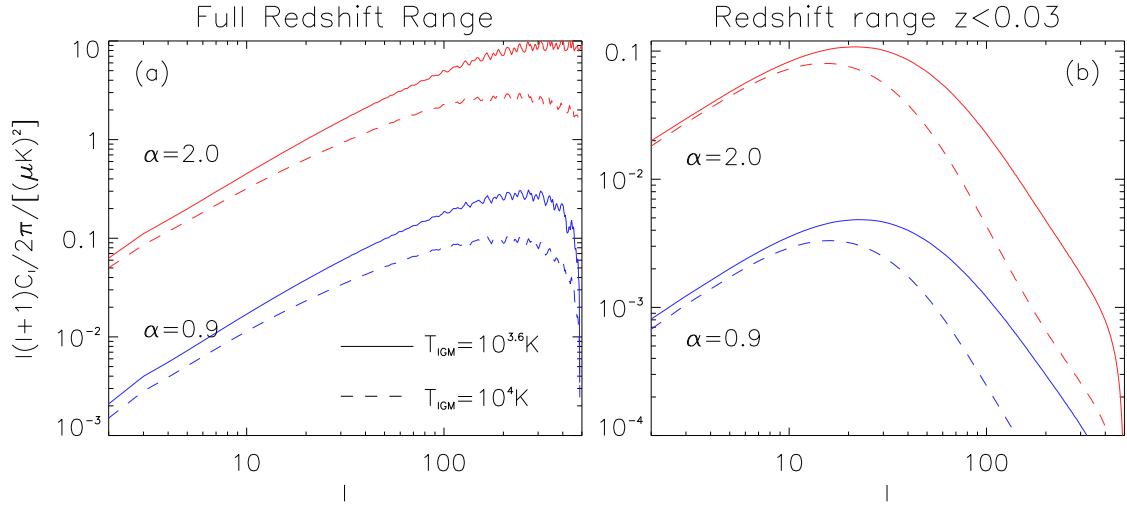}
\vspace*{-1cm}
\caption{\small  
Power spectra of the tSZ due to the WHIM for different model parameters.
As indicated the spectra in panel (a) includes the 
contribution at all redshifts, while the spectra in panel (b) only includes 
the contribution up to $z_{\rm up}=0.03$. Solid and dashed lines
correspond to $T_{\rm IGM}=10^{3.6}$~K and $10^{4}$~K, respectively; 
neighboring graphs, given in the same color, correspond to the same value 
of the equation of state parameter $\alpha$, in decreasing amplitude from
top to bottom.
}
\label{fig5}
\end{figure}

To compute the tSZ effect generated by the free electrons in the WHIM, we need
to specify the equation of state. For shock-heated gas, the phase diagrams 
obtained by \citet{kang05} from numerical simulations can be fitted by
$\log_{10}(T_{\rm e}/10^8K)=-2[\log_{10}(3.5+x^b)]^{-1}$ with $b=\alpha+x^{-1}$ and 
$x=n_{\rm e}/\bar{n}_{\rm B}$ the electron density in units of the mean baryon density.  
In addition to the two parameters $T_{\rm IGM}$ and $\alpha$, 
WHIM temperature anisotropies depend on $\sigma_8$, the amplitude of the
matter power spectrum on a sphere of $8h^{-1}$~Mpc, and on the fraction of
free electrons in the filaments. In Figure~\ref{fig5} we present the radiation power 
spectra of the WHIM temperature anisotropies for different model parameters;
we assume that all the baryons in the WHIM are ionized. In these models
we fixed the Planck measured value of $\sigma_8=0.83$ \citep{cpp16}.
This corresponds to a WHIM baryon fraction of 43\% to 48\% 
for $T_{\rm IGM}=10^{3.6},\; 10^{4}$K, respectively. The spectra in panel 
(a) of Figure~\ref{fig5} represent the full contribution, obtained by
integrating up to redshift $z_{\rm up}=3$. The contribution of filaments
from earlier times is negligible. Spectra are presented in pairs,
corresponding to $T_{\rm IGM}=10^{3.6}$~K (solid) and $10^4$~K (dashed lines),
respectively. From top to bottom, curves correspond to a equation of state 
parameter $\alpha=2.0, 0.9$, respectively. 
In panel (b) of Figure~\ref{fig5} the power spectra follow the same convention. 
In this case the integration is restricted to $z_{\rm up}<0.03$, similar to the
scale probed by our template. 

Our formalism allows us to predict the cross-correlation of the spatial 
distribution of the WHIM with the tSZ distortion generated by it.
As indicated in section~\ref{sec:templates}, the matter template describes 
the matter distribution in the mildly non-linear regime. If the ionized gas
follows the matter and its equation of state is polytropic, 
$T_{\rm e}\propto n_{\rm e}^{(\gamma-1)}$, then the pressure 
profile will be $P_{\rm e}\propto n_{\rm e}^\gamma$, where $\gamma=1$
corresponds to isothermal gas, characteristic of the shocked regions
and $\gamma=5/3$ to adiabatic monoatomic gas. From the template
$M\propto n_e$ we have constructed pressure templates with $\gamma=(1,5/3)$.
We verified that the cross-correlation at zero lag varied by less than 20\% 
for adiabatic indices in the above range,
indicating that if the template describes the distribution of the
ionized WHIM, then it will also trace its tSZ contribution. Then, 
if a matter template, normalized to zero mean and unit variance,
traces the tSZ, the cross-correlation would be 
$\langle TM\rangle (0)=\sigma_{\rm WHIM}$ where
$\sigma^2_{\rm WHIM}=\sum (2\ell+1)C_\ell^{\rm WHIM}/4\pi$; therefore, the
results presented in Figure~\ref{fig5} give the expected correlation
for each set of model parameters. Additionally, we need to consider
that the actual realization of the WHIM in the sky is not exactly
the mean power spectrum since our template is just one realization 
of the sky. This gives rise to an uncertainty known as cosmic variance 
and given by $\epsilon_{\rm CV}=\sum\sqrt{2(2\ell+1)}C_\ell/4\pi$.

\begin{figure}[t]
\centering
\plotone{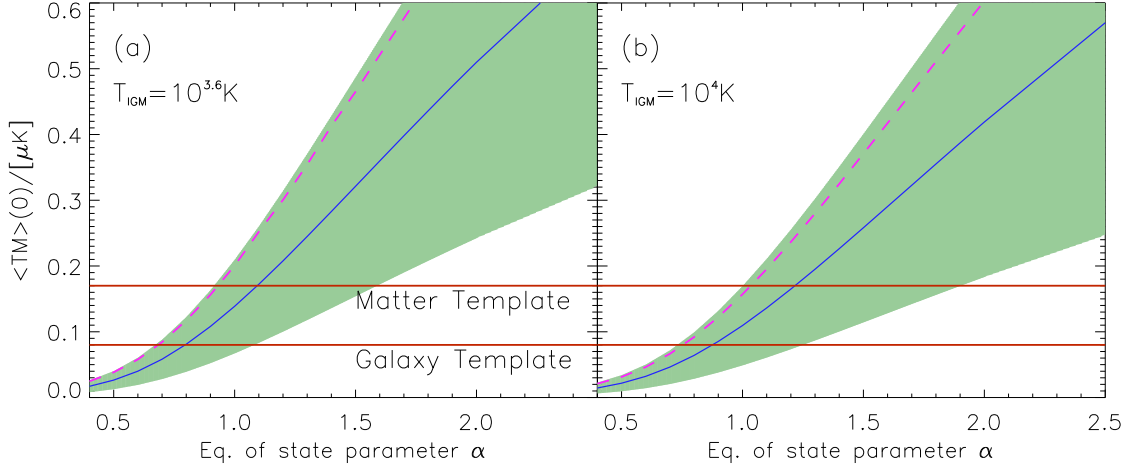}
\caption{\small 
Comparison of the measured and the predicted amplitude of the cross-correlation 
of the matter template that traces the matter distribution out to
$z_{\rm up}=0.03$ and the tSZ anisotropy due to the WHIM. The horizontal 
red lines give the upper limit of the measured correlation at zero lag at
the 95\% confidence levels derived from the matter template and 
the galaxy templates. The blue solid line and the
shaded green area represent the expected cross-correlation and the $1\sigma$
cosmic variance uncertainty for different model parameters, considering a maximum 
redshift $z_{\rm up}=0.03$ for the matter density field traced by the template. The 
dashed violet line represents the predicted cross-correlation for $z_{\rm up}=0.05$.
In panel (a) and (b) the IGM temperature is $T_{\rm IGM}=10^{3.6}$K and $10^{4}$K, 
respectively. 
}
\label{fig6}
\end{figure}

In Figure~\ref{fig6} we present the amplitude of the cross-correlation
at zero-lag, computed using the upper limit of integration $z_{\rm up}=0.03$,
similar to the depth of our template with the upper limits 
obtained from the data. In panels (a) and (b) of Figure~\ref{fig6}
the model corresponds to $T_{\rm IGM}=10^{3.6}, 10^{4}$~K, respectively. 
Plots follow the same convention; the thick solid blue lines represent the 
theoretical prediction and the surrounding shaded green area the $1\sigma$ cosmic
variance uncertainty. The horizontal (red) lines correspond to the
95\% confidence limit given by the measured cross-correlation with the
matter and the galaxy templates.
The dashed (violet) line represents the correlation for
an upper limit of integration $z_{\rm up}=0.05$ to show that by increasing 
the depth of our template we could probe the model very effectively. Since
our error bars are dominated by the random alignments of our filaments with
large scale structures in the CMB data, they will not increase
significantly if our template was to trace the matter distribution
out to, for example, that redshift but the signal will
rise by almost a 50\%. If we were not to have a detection
then the constraints on model parameters would be much stronger. 

Using a grid of models in the range $\sigma_8=[0.7,1]$, $\alpha=[1,4]$ and
$\log_{10}(T_{\rm IGM})=[3.4,4.2]$ we have verified that the power spectra 
scales with cosmological and model parameters as
$\sigma_{\rm WHIM}\propto (\log T_{\rm IGM})^{-1.4\pm 0.2}
\alpha^{1.7\pm 0.2} \sigma_8^{1.3\pm 0.1} \Omega_{\rm B}$.
Since $\sigma_8$ and $\Omega_{\rm B}$ are fixed by observations 
the results of Figure~\ref{fig6} show that the equation of state parameter
$\alpha$ is the most important parameter of the model. In the observationally
allowed range, the variation of the model prediction with $T_{\rm IGM}$ is
rather weak.  Including cosmic variance and considering the upper limit
at the 95\% confidence level of the matter
template, $\alpha\lesssim 1.6$ and $\lesssim 1.9$ for 
$T_{\rm IGM}=10^{3.6}$~K and $10^{4}$~K, respectively. The bound is 
much more stringent when we impose the upper limit derived from the
galaxy template: $\langle TM_g\rangle(0)<0.08~\mu$K:
$\alpha\lesssim 1.1$ and $\alpha\lesssim 1.3$, respectively for the two values 
of $T_{\rm IGM}$ considered. 

The upper bound in $\alpha$ can be translated into a constraint on the 
temperature of the gas at different densities using the equation of state derived
from the results of Kang et al (2005). Figure~\ref{fig7} defines the 
range of the temperature of the shocked-heated WHIM as a function of its 
density contrast. The solid line corresponds to $\alpha=1.0$ and 
the dashed lines correspond to the 95\% confidence level
upper limit from the matter template, $\alpha=1.9$ (upper line) and 
the same limit from the galaxy template, $\alpha=1.3$ (lower line).
The shaded green area shows the region excluded by the more strict upper limit 
from the galaxy template that is still allowed by the matter template. For 
comparison we also plot the same temperature function for the numerical
results of \cite{cen06}. For instance, at overdensity $x=1$ the temperature 
of the electron gas is $T_e\simeq 10^{4.8}$K, while for $x=100$ it is 
$T_e\lesssim 10^{7.5}$K. These temperature ranges are
in agreement with the results of simulations (see \citet{kang05,cen06,bertone10}). 
Still, at $T_{\rm IGM}=10^{3.6}-10^4$K
the fraction of baryons in the WHIM medium varies from 43\% to 48\%, providing
enough room to accommodate the missing baryon fraction in the local
Universe.

\section{Conclusions}

We have searched for the tSZ signature of the WHIM in Planck Nominal maps. 
To that purpose we have cleaned the frequency maps of Planck first data release 
from foreground emissions. These maps were correlated with a template
reconstructed from the observed galaxy distribution in the 2MRS 
that describes the matter density field out to $\approx 90 h^{-1}$~Mpc that
traces the large-scale filamentary distribution of the WHIM
and with a galaxy template that traces the hot circumgalactic halos.
Our results showed no evidence of a WHIM-induced tSZ temperature anisotropy
to be present at any significant level or of
any detectable contribution from virialized gas within low-mass halos.
We set upper limits of $<0.08~\mu$K  
and $<0.17~\mu$K at $95\%$ C.L. using the results from the 
galaxy and matter templates, respectively. We translated these upper 
limits into constraints on the properties of the WHIM
using the model described in \citet{suarez13b}.
We found that the temperature of the shock-heated WHIM gas was in the range
$T=10^{4.5}-10^{7.5}$K at overdensities in the range $1-100$. 
The cross-correlation of the matter template and the Planck CMB
data goes to zero at $\theta\sim 6^\circ$. At the median depth of our template,
this angular scale would be compatible with filaments having an averaged
projected size of $6-8h^{-1}$Mpc. The size of filaments and the temperature
of their gas is in excellent agreement with the scale and temperature range 
found by Van Waerbeke et al (2014), indicating that the matter template could be 
tracing the WHIM distribution even though it is not deep enough to produce a 
statistically significant signal. 

\begin{figure}[t]
\centering
\includegraphics[width=12cm]{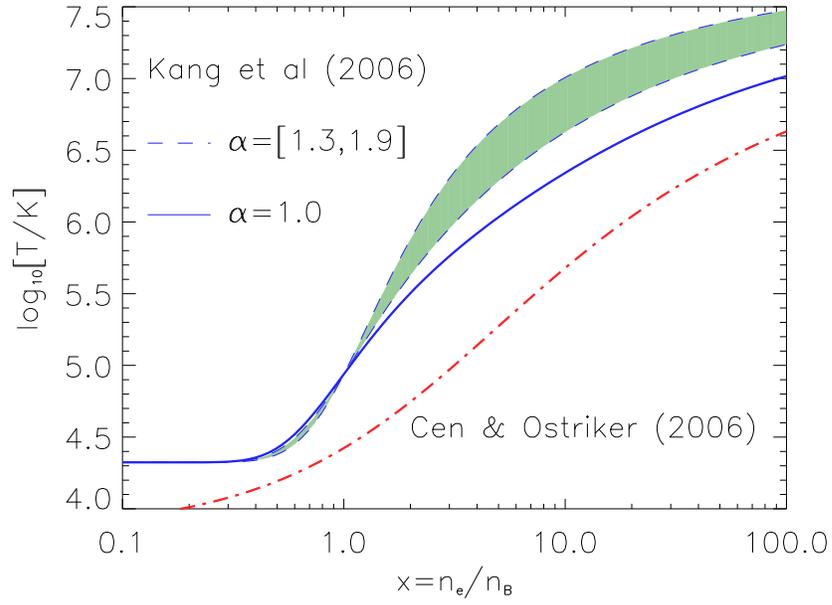}
\caption{\small The blue solid line represents the 
temperature of the shock-heated WHIM for different overdensities
in the Kang et al (2005) numerical fit for an equation of state
parameter $\alpha=1$. The blue dashed lines correspond
to $\alpha=1.3$ (bottom) and $\alpha=1.9$ (top), the 95\% upper limits
from the galaxy and matter templates, respectively. The shaded green 
area shows the region excluded by the more strict constraint of the
galaxy template that is allowed by the matter template. 
For comparison, the dot-dashed red line represents the 
temperature-overdensity relation from \citet{cen06}.
}
\label{fig7}
\end{figure}

The search for the missing baryons is not restricted to the CMB. Evidence of 
the distribution of this matter component has been obtained via UV-absorption 
line studies with FUSE and HST-COS (Shull et al. 2012) but these observations 
probe the coldest fraction of these baryons. Our results imply that
most baryons would have temperatures $T\ge 10^{4.5}$K, in agreement
with hydrodynamical simulations. This phase could be detected through 
highly ionized C, N, O, Ne, and Fe ions. In this respect, the proposed
Athena X-ray mission (Kaastra et al 2013) could provide very useful
information about the WHIM parameters. With WHIM gas temperatures
in the range $T=10^{4.5}-10^{7.5}$K, the filaments in our model can accommodate
up to 48\% of all the baryons. Since the tSZ effect is proportional
to the electron pressure, if X-ray observations were to find that the 
average WHIM is larger than the limits given in Figure~\ref{fig7} then,
to make compatible the tSZ upper limit with a higher WHIM temperature,
the baryon fraction stored in filaments must decrease in a proportion similar 
to the temperature increment. Then, by combining observations
of the WHIM at different wavelengths and different redshifts, we
will measure better the physical properties of the WHIM, provide
better constraints on the model parameters, improve our understanding
of the physical processes undergone by the baryons during the formation
of galaxies and the emergence of large scale structure, and 
could separate the contribution of gas in unresolved halos from
that of the WHIM.

\vspace*{1cm}
We acknowledge the use of data from the Planck/ESA mission, downloaded 
from the Planck Legacy Archive. RGS acknowledges financial support from 
the Spanish grant CONSOLIDER EPI (CSD2010-00064).
FAB acknowledges financial support from the Spanish
Ministerio de Educaci\'on y Ciencia (grant FIS2012-30926) and
to the ``Programa de Profesores Visitantes Severo Ochoa'' of the Instituto
de Astrof\1sica de Canarias. FSK is a Karl-Schwarzschild fellow at the AIP.

\clearpage
\pagestyle{plain}

\clearpage
\pagestyle{plain}

\clearpage
\pagestyle{plain}

\clearpage
\pagestyle{plain}

\clearpage
\vspace*{.5cm}
\pagestyle{plain}

\clearpage
\pagestyle{plain}

\clearpage
\pagestyle{plain}


\begin{thebibliography}{99}

\bibitem[Anderson \& Bregman (2011)]{anderson11}{
Anderson, M.~E. \& Bregman, J.~N. 2011, ApJ, 737, 22}

\bibitem[Atrio-Barandela \& M\"ucket (2006)]{atrio06}{
Atrio-Barandela, F., \& M\"ucket, J.~P. 2006, ApJ, 643, 1}

\bibitem[Atrio-Barandela et al.(2008)]{atrio08}{
Atrio-Barandela, F., M\"ucket, J.~P., \& G\'enova-Santos, R.,
2008, ApJ, 674, L61}

\bibitem[Bertone et al.(2010)]{bertone10} Bertone, S., Schaye, 
J., Dalla Vecchia, C., et al.\ 2010, \mnras, 407, 544 

\bibitem[Buote et al.(2009)]{buote09} Buote, D.~A., Zappacosta, 
L., Fang, T., et al.\ 2009, \apj, 695, 1351 

\bibitem[Cen \& Ostriker (1999)]{cen99}{Cen, R., \& Ostriker, J.P. 1999, ApJ, 519, L109}

\bibitem[Cen \& Ostriker (2006)]{cen06}{Cen, R., \& Ostriker, J.P. 2006, ApJ, 650, 560}

\bibitem[Dai et al.(2012)]{dai12}{
Dai, X.  Anderson, M.~E., Bregman, J.~N. \& Miller, J.M.  2012, ApJ, 755, 107}

\bibitem[Danforth \& Shull(2008)]{danforth08} Danforth, C.~W., \& Shull, J.~M.\ 2008, \apj, 679, 194 

\bibitem[Dave et al.(2001)]{dave01}{Dav\'e, R., et al. 2001, ApJ, 552, 473}

\bibitem[Fukugita \& Peebles (2004)]{fukugita04}{
Fukugita, M., \&  Peebles, P.J.E. 2004, ApJ, 616, 643}

\bibitem[G{\'e}nova-Santos et al.(2005)]{genova05} G{\'e}nova-Santos, 
R., Rubi{\~n}o-Mart{\'{\i}}n, J.~A., Rebolo, R., et al.\ 2005, \mnras, 363, 79 

\bibitem[G\'enova-Santos et al.(2009)]{genova09}{G\'enova-Santos, R., 
Atrio-Barandela, F., M\"ucket, J.~P., \&  Klar, J.S., 2009, ApJ, 700, 447}

\bibitem[G\'enova-Santos et al.(2013)]{genova13}{
G\'enova-Santos, R., Suarez-Vel\'asquez, I., Atrio-Barandela, F. \& M\"ucket, J.~P.
2013, MNRAS, 432, 2480}

\bibitem[Gorski et al.(2005)]{healpix}{
Gorski, K. et al 2005, Ap.J., 622, 759}

\bibitem[Hern\'andez-Monteagudo et al.(2004)]{hernandez04}{
Hern\'andez-Monteagudo, C., G\'enova-Santos, R., \& Atrio-Barandela, F., 
2004, ApJ, 613, L89}

\bibitem[He{\ss} et al.(2013)]{hess13} He{\ss}, S., Kitaura, 
F.-S., \& Gottl{\"o}ber, S.\ 2013, \mnras, 435, 2065 

\bibitem[Huchra et al.(2012)]{huchra12} Huchra, J.~P., Macri, 
L.~M., Masters, K.~L., et al.\ 2012, \apjs, 199, 26 

\bibitem[Kaastra et al.(2013)]{kaastra13} Kaastra, J., Finoguenov, A., Nicastro, F.
et al. 2013, ArXiv:1306.2324

\bibitem[Kang et al.(2005)]{kang05} Kang, H., Ryu, D., Cen, 
R., \& Song, D.\ 2005, \apj, 620, 21 

\bibitem[Kitaura (2012a)]{kitaura12}{
Kitaura, F.-S. (2012a) MNRAS, 420, 2737}

\bibitem[Kitaura et al.(2012b)]{kitaura12b}{
Kitaura, F.-S., Erdogdu, P., Nuza, S. E., Khalatyan, A., Angulo, R. E., Hoffman, 
Y., Gottl\"ober, S., 2012b, MNRAS, 427, 35}

\bibitem[Klar \& M\"ucket(2010)]{klar10} Klar, J.~S., \& M\"ucket, 
J.~P.\ 2010, \aap, 522, 114 

\bibitem[Molnar \& Birkinshaw(1998)]{molnar98} Molnar, S.~M., 
\& Birkinshaw, M.\ 1998, \apj, 497, 1 

\bibitem[Nicastro et al.(2005)]{nicastro05} Nicastro, F., Mathur, S., 
Elvis, M., et al.\ 2005, \apj, 629, 700

\bibitem[Planck Collaboration (2013a)]{pip8}{Planck Collaboration.
Planck Intermediate Results VIII, 2013a, A\& A, 550, 134}

\bibitem[Planck Collaboration (2013b)]{pip11}{Planck Collaboration.
Planck Intermediate Results XI, 2013b, A\& A, 557, 52}

\bibitem[Planck explanatory supplement (2013)]{pes} Planck Collaboration. 
Planck explanatory supplement. First release v1.00 \ 2013
{\tt http://wiki.cosmos.esa.int/planckpla/index.php/Main\_Page}

\bibitem[Planck Collaboration et al.(2014a)]{cpp1} Planck Collaboration. 
Planck 2013 Results I, 2014a, \aap, 571, 1 

\bibitem[Planck Collaboration et al.(2014b)]{cpp6} Planck Collaboration. 
Planck 2013 Results VI, 2014b, \aap, 571, 6 

\bibitem[Planck Collaboration et al.(2014c)]{cpp11} Planck Collaboration. 
Planck 2013 Results XI, 2014c, \aap, 571, 11

\bibitem[Planck Collaboration et al.(2014d)]{cpp12} Planck Collaboration. 
Planck 2013 Results XII, 2014d, \aap, 571, 12

\bibitem[Planck Collaboration et al.(2014e)]{cpp13} Planck Collaboration. 
Planck 2013 Results XIII, 2014e, \aap, 571, 13

\bibitem[Planck Collaboration (2014)f]{cpp16}{Planck Collaboration.
Planck 2013 Results XVI, 2014f, A\& A, 571, 16}

\bibitem[Planck Collaboration et al.(2014g)]{cpp21} Planck Collaboration. 
Planck 2013 Results XXI, 2014g, \aap, 571, 21 

\bibitem[Rauch et al.(1997)]{rauch97}{
Rauch, M., Miralda-Escud\'e, J. \& Sargent, W. L. 1997, ApJ, 489, 7}

\bibitem[Suarez-Vel\'asquez et al.(2013a)]{suarez13a}{
Suarez-Vel\'asquez, I., Kitaura, F.-S., Atrio-Barandela, F. \& M\"ucket, J. P.,
2013a, ApJ, 769, 25}

\bibitem[Suarez-Vel\'asquez, M\"ucket \& Atrio-Barandela (2013b)]{suarez13b}{
Suarez-Vel\'asquez, I.~F., M\"ucket, J.~P., \& Atrio-Barandela, F. 2013b, MNRAS, 431, 342}

\bibitem[Shull et al.(2012)]{shull12}{
Shull, J.~M., Smith, B.~D., \& Danforth, D.~W. 2012, ApJ, 759, 23}

\bibitem[Sunyaev \& Zel'dovich (1970)]{sunyaev70}{
Sunyaev, R.~A., \& Zel'dovich, Y.~B. 1970, ApSS, 7, 3}

\bibitem[Tittley \& Meiksin (2007)]{tittley07}{
Tittley E., Meiksin A., 2007, MNRAS, 380, 1369}

\bibitem[Tripp et al.(2008)]{tripp08} Tripp, T.~M., Sembach, K.~R., Bowen, D.~V., et al.\ 2008, \apjs, 177, 39 

\bibitem[Ursino et al.(2014)]{ursino14}{
Ursino, E., Galeazzi, M. \& Huffenberger, K. 2014, ApJ, 789, 55}

\bibitem[Van Waerbeke et al.(2014)]{van_waerbeke14}{
Van Waerbeke, L., Hinshaw, G. \& Murray, N. 2014, PRD, 89, 023508}

\bibitem[Werner et al.(2008)]{werner08} Werner, N., Finoguenov, A., Kaastra, J.~S., et al.\ 2008, \aap, 482, L29 

\bibitem[Weinberg et al.(1997)]{weinberg97}{
Weinberg, D. H., Miralda-Escud\'e, J., Hernquist, L. \& Katz, N. 1997, ApJ, 490, 564}

\bibitem[Williams et al.(2013)]{williams13} Williams, R.~J., 
Mulchaey, J.~S., \& Kollmeier, J.~A.\ 2013, \apjl, 762, LL10 

\bibitem[Yao et al.(2012)]{yao12}{
Yao, Y., Shull, J. M., Wang, Q. D. \& Cash, W. 2012, ApJ, 746, 166}

\end{thebibliography}
\end{document}